\documentclass[10pt,letterpaper,twocolumn]{article} 

\usepackage{ol2}
\usepackage[draft]{hyperref}
\usepackage{amsmath}

\begin{document}

\twocolumn[ 

\title{Large negative lateral shifts due to negative refraction}


\author{Jessica Benedicto,$^{1,2}$ R\'emi Poll\`es,$^{1,2}$ Antoine Moreau,$^{1,2,*}$ Emmanuel Centeno,$^{1,2}$}

\address{
$^1$ Clermont Universit\'e, Universit\'e Blaise Pascal, LASMEA, BP 10448, F-63000 Clermont-Ferrand\\
$^2$ CNRS, UMR 6602, LASMEA, F-63177 Aubi\`ere, France\\
$^*$ Corresponding author: antoine.moreau@univ-bpclermont.fr
}

\begin{abstract}When a thin structure in which negative refraction occurs (a metallo-dielectric or a photonic crystal) is illuminated by a beam, the reflected and transmitted beam can undergo a large negative lateral shift. This phenomenon can be seen as an interferential enhancement of the geometrical shift and can be considered as a signature of negative refraction.\end{abstract}


 ] 

It has been recently shown that when a beam is illuminating a slab
presenting negative permittivity and permeability the reflected and
transmitted beam could undergo a large negative lateral
shift\cite{wang05}. It has then been suggested that some of these
lateral shifts are linked to the negative refraction which occurs when
light enters the structure\cite{moreau07,moreau08}. In the present
letter, we show for the first time that a large lateral shift
explicitely due to negative refraction can be obtained using a very
thin slab of photonic crystal, thus confirming the previous
assumption. Here we have not used metamaterials designed
to present negative permittivity and permeability as in\cite{wang05},
which definitely shows that negative refraction alone is responsible
for the negative shift.

In previous works, large lateral shifts have already been obtained
using a photonic crystal, but they were either positive\cite{felbacq03,paul08} or negative
but due to a guided mode excited by a grating\cite{he06}.


The lateral shift associated to negative refraction has already been
proposed as a signature of that phenomenon\cite{kong} in a case where
the thickness of the structure is larger than the waist of the
incident beam. The lateral shift is in that case purely geometrical.
We show in this paper that the large negative lateral shift is finally
an interferential enhancement of the geometrical lateral shift. It 
occurs in very thin structures compared to the waist, and
it can be used to characterize negative refraction.

In order to get a physical picture of the phenomenon, let us begin by
considering a {\it lossless} metallo-dielectric multilayer. Such a
structure can be considered as a 1D photonic crystal.
It is now well-known that negative refraction may occur when TM
polarized light enters such a structure\cite{scalora07}. The
transparency of that type of structure is due to the excitation of
coupled resonances : either coupled guided modes supported by the
metallic slabs, or Fabry-Perot resonances of the cavity constituted by
the dielectric layers. Here we consider the second case, when the
field is propagative in the dielectric layer.

Negative refraction occurs when the dielectric layers are thin
and when their relative permittivity is high enough, 
so that the field is more localized in the metal. 
We have thus chosen $\epsilon_m=-4.43$, corresponding to the real part of
the permittivity of silver at $400 \; nm$ and $\epsilon_d=5.3$ so that
$\epsilon_d>|\epsilon_m|$. The thickness of the dielectric
(resp. metallic) layers is $h_d=0.0979\lambda$ (resp. $h_m=0.0975\lambda$).

The component of the Poynting vector parallel to the interfaces of the
metallo-dielectric structure, $P_x$, controls the direction in which light will
propagate in the crystal. When that quantity, averaged on a period of
the structure, is negative, then negative refraction
occurs\cite{benedicto11}. Inside a layer, it can be written
\begin{equation}
P_x=\frac{n\,\sin(\theta)}{2c\epsilon_0\epsilon_r}\,|H|^2
\end{equation}
where $n$ is the refractive index of the above medium in which the
incident beam propagates with an incidence angle $\theta$ and
$\epsilon_r$ is the relative permittivity of the medium in which the
vector is calculated.

\begin{figure}[htb]
\includegraphics[width=8.3cm]{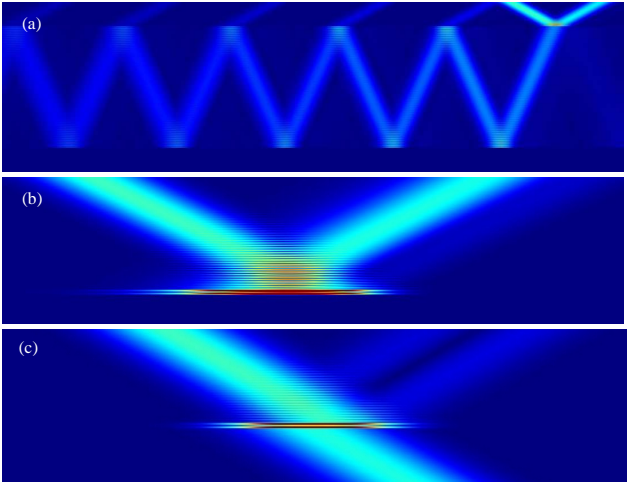}
  \caption{Metallo-dielectric structure illuminated by a TM polarized
    gaussian beam with a $61.7^\circ$ angle of incidence, $10\lambda$
    waist in two cases (a) when the structure contains 500 periods, in
    the geometrical regime (b) when the structure is only 10 periods
    thick, in the interferential regime for which a large negative
    lateral shift occurs. (c) When it is present, the transmitted beam
 too undergoes a negative lateral shift (structure with 5 periods,
 $0.997\;nm$ thick, $60.65^\circ$). This shift is accurately given, for
 large beams, by Artmann's formula (\ref{artmann}) considering the phase of 
the {\it transmission} coefficient.
\label{fig:un}}
\end{figure}

Using a code based on the scattering matrix method which is freely available\cite{krayzel10} we have
considered a beam illuminating a thick structure containing 500
periods of the photonic crystal. In that case the waist of the beam
($10\lambda$) is small compared to the thickness of the structure (around $100\lambda$) :
this is the geometrical regime. Figure \ref{fig:un} (a) shows a
situation which is very similar to\cite{kong}, except that there are
many reflected beams inside the structure. This is explained by the
high reflectivity of the interfaces between the photonic crystal and
the outer media : above the structure we have chosen a dielectric with
$\epsilon_r=6.7$ and under the photonic crystal there is air. Given
the incidence angle of the incoming beam, total reflexion occurs at
the lowest interface.

When the thickness of the overall structure is reduced, all the beams
reflected inside the structure interfere. If they interfere
constructively, a Fabry-Perot resonance occurs in the photonic crystal
slab. Since we are not at normal incidence, that resonance is actually
a leaky mode\cite{moreau08} and it is backward because the beams
inside the slab propagate towards the left because of negative
refraction. The excitation of this leaky mode leads to enhance 
the lateral shift by an interferential effect.  Figure
\ref{fig:un} (b) shows such a situation for an incidence angle of
$61.7^\circ$. Despite the very small thickness of the photonic crystal (10
periods for a overall thickness of $2\lambda$), the lateral shift
($14.8\lambda$) is much larger than the waist of the beam and it cannot
be explained by a geometrical lateral shift.

\begin{figure}[h]
\includegraphics[width=8.3cm]{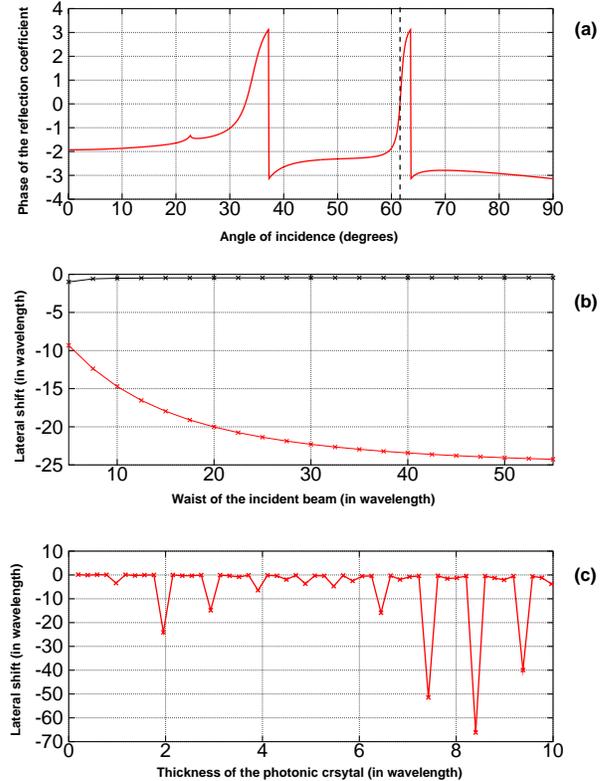}
  \caption{(a) Phase of the reflection coefficient for the 10 periods structure
considered fig. \ref{fig:un} showing that the phase presents a positive derivative,
characteristics of the negative lateral shift (b) Lateral shift as a function 
of the waist for an incidence angle of $61.7^\circ$ corresponding
to a swift variation of the phase (red curve) and for $58^\circ$ (black curve)
 leading to a smaller shift.(c) Lateral shift  as a function of the thickness.
\label{fig:deux}}
\end{figure}

Such an interferential effect does not occur for any incidence
angle. The angles for which the reflected beam is shifted
can be found by considering the phase of the reflection
coefficient of the photonic crystal slab. The lateral shift of a
reflected beam is actually given by Artmann's formula\cite{tamir86,felbacq03}
\begin{equation}
 \Delta = - \frac{1}{nk_0 \cos \theta} \frac{\partial \phi}{\partial
   \theta}, \label{artmann}
\end{equation}
where $\theta$ is the angle of incidence, $n$ the index of the upper
medium, $k_0 = \frac{2 \pi}{\lambda}$ the wavenumber in vacuum, and
$\phi$ the phase of the reflection coefficient. That shift is the
asymptotic shift, reached when the waist of the incident beam is large
enough\cite{krayzel10}. Artmann's formula indicates here that a backward
leaky
mode can be excited for angles for which the phase presents a quick
{\em positive} variation (due to the presence of a pole\cite{moreau08}).
Figure \ref{fig:deux} (a) shows the phase of the
reflection coefficient of the thin photonic crystal slab and how we
have chosen the angle of incidence.

But it should be stressed that even though the lateral shift shown
figure \ref{fig:un}(b) would allow an easy measurement, the waist of
the incoming beam is not large enough to reach the asymptotic regime -
so that the shift is smaller than what is given by (\ref{artmann}). That
is however not a problem, because when Artmann's formula is valid the
waist is in general so large than the lateral shift is harder to
detect\cite{shad03}. We are confident that such a lateral shift can easily be measured
because positive lateral shifts have already been measured either for
large lateral shifts\cite{pillon05} or even Goos-H\"anchen
shifts\cite{bretenaker} which are much smaller.

However, for a negative lateral shift to be
considered as a sign of negative refraction, one has to be sure that it
is not due to the excitation of a leaky surface
mode\cite{moreau07,laroche08}. The only way to make sure a leaky cavity
mode has been excited is to study the dependence of the negative lateral
shift on the thickness of the structure (see figure \ref{fig:deux} (c)).
Our structure being a photonic crystal, it is not possible to make the
thickness vary continuously. It is then not possible to see a
periodical variation of the shift (as could be expected) but a beat
between the periodical shift and the sampling rate as shown in the
figure. Such a behaviour is characteristic of a cavity resonance.

Since losses shorten the propagation length of light in the structure,
they always reduce the number of interfering beams and finally the 
negative lateral shift. But the interferential enhancement can be
expected to occur as long as the propagation length is several times
larger than the actual thickness of the structure. This can be easily
verified by considering a slightly lossy metal in the above structures,
but not when the losses are realistic. In that case, the propagation
length is about a few periods of the structure\cite{scalora07}.

\begin{figure}[h]
\includegraphics[width=8.3cm]{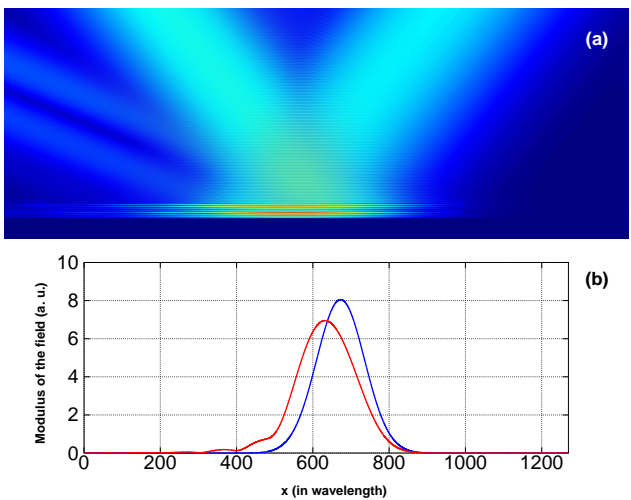}
\caption{(a) Gaussian beam illumating a thin slab (7 periods) of a
  photonic crystal with an incidence angle of $40^\circ$ and a waist of
  $89\lambda$ (b) profile of the reflected $H$ field showing how it is
  slightly distorted and shifted towards the left. The field plotted here
is the zero-order only. The $-1$ diffraction order can be seen 
on the left of the picture - its profile is due partly to the incoming beam
and partly to the resonance inside the structure.\label{fig:trois}}
\end{figure}

We have then chosen in the literature\cite{gralak00} a bidimensional photonic
crystal which produces negative refraction and which, since it is
purely dielectric, does not suffer from losses. The crystal is made of
rectangular rods (width of $0.75\lambda$, height of $0.96\lambda$, permittivity $\epsilon_r=9$)
periodically arranged in air - with an horizontal period of
$1.27\lambda$ and a vertical period of $2.22\lambda$. In order to
obtain an enhancement of the lateral shift, light must be, as much as
possible, reflected inside the structure. It is not possible to hinder
light from leaking in the above medium here, so that a good solution
is always to have light completely reflected at the lower
interface. In general, that can be obtained using total reflection
like in the previous case, but it is not possible here because the
above medium is air. Using a back reflector (a Bragg mirror or a
metallic mirror) is a possibility. Here for simplicity we have chosen
to place the photonic crystal on a perfect conducting
reflector. Figure \ref{fig:trois} shows the simulation we have made
using a Fourier modal method. Despite the fact that the reflection
coefficient between the crystal and air has not been particularly
optimized, a Fabry-Perot resonance of the slab can be clearly seen,
leading to a $-36.6\lambda$ lateral shift of the reflected beam (with
a waist of $89\lambda$).

To summarize, for a structure producing negative refraction it is
possible to obtain a large negative lateral shift provided (i) there
is a high reflectivity for light inside the structure, which can be
achieve by using either a back reflector or an impedance mismatch
between the photonic crystal and the outside media (ii) the thickness
of the structure is small compared to the waist of the incident beam,
so that the different beams inside the structure can interfere (iii) a
correct angle is chosen, so that the interferences inside the structure
are constructive and a resonance is excited. This phenomenon takes
place even if the waist of the incident beam is not large enough so
that Artmann's formula, which is widely used by the community to
characterize lateral shifts, is not valid. And in general, it is
easier to measure it when the asymptotic regime is not reached.

In conclusion, we have shown in the present paper that photonic crystal slabs could
support backward leaky modes which are responsible for a large negative
lateral shift of the reflected beam. These large negative lateral shifts
can be used to characterize negative refraction because they can be seen
as an interferential enhancement of the geometrical negative lateral
shift\cite{kong}. It must be underlined that by nature these shifts occur when the thickness of the
structure is small compared to the waist of the incoming beam.

\vspace*{-0.1cm}

\end{document}